\documentclass[conference]{IEEEtran}
\usepackage{amsmath,amssymb,epsfig}
\usepackage{cite,url}
\begin{document}
\title{Asymmetric Interference Cancellation for 5G Non-Public Network with Uplink-Downlink Spectrum Sharing}
\author{\IEEEauthorblockN{Peiming~Li\IEEEauthorrefmark{1}\IEEEauthorrefmark{2},  Lifeng~Xie\IEEEauthorrefmark{1}\IEEEauthorrefmark{2}, Jianping Yao\IEEEauthorrefmark{1}, Jie~Xu\IEEEauthorrefmark{2}, Shuguang Cui\IEEEauthorrefmark{2}, and Ping Zhang\IEEEauthorrefmark{3}}
\IEEEauthorblockA{\IEEEauthorrefmark{1}
%School of Information Engineering
SIE, Guangdong University of Technology, Guangzhou 510006, China}
\IEEEauthorblockA{\IEEEauthorrefmark{2}
FNii and SSE, The Chinese University of Hong Kong, Shenzhen, Shenzhen 518172, China}
\IEEEauthorblockA{\IEEEauthorrefmark{3}
%State Key Laboratory of Networking and Switching Technology,
%School of Information and Communication Engineering
SKLNST and SICE, Beijing University of Posts and Telecommunications, Beijing 100876, China}
E-mail:
peiminglee@outlook.com, lifengxie@mail2.gdut.edu.cn, yaojp@gdut.edu.cn, \\\{xujie, shuguangcui\}@cuhk.edu.cn, pzhang@bupt.edu.cn
}
\maketitle
\newcommand\blfootnote[1]{
\begingroup
\renewcommand\thefootnote{}\footnote{#1}
\addtocounter{footnote}{-1}
\endgroup
}
\blfootnote{J. Xu is the corresponding author.}
\begin{abstract}
Different from public 4G/5G networks that are dominated by downlink traffic, emerging 5G non-public networks (NPNs) need to support significant uplink traffic to enable emerging applications such as industrial Internet of things (IIoT). The uplink-and-downlink spectrum sharing is becoming a viable solution to enhance the uplink throughput of NPNs, which allows NPNs to perform the uplink transmission over the time-frequency resources configured for downlink transmission in coexisting public networks. To deal with the severe interference from the downlink public base station (BS) transmitter to the coexisting uplink non-public BS receiver, we propose an adaptive asymmetric successive interference cancellation (SIC) approach, in which the non-public BS is enabled to have the capability of decoding the downlink signals transmitted from the public BS and cancelling them for interference mitigation. In particular, this paper studies a basic uplink-and-downlink spectrum sharing scenario when an uplink non-public BS and a downlink public BS coexist in the same area, each communicating with multiple users via orthogonal frequency-division multiple access (OFDMA). Under this setup, we aim to maximize the common uplink throughput of all non-public users, under the condition that the downlink throughput of each public user is above a certain threshold. The decision variables include the subcarrier allocation and user scheduling for both non-public and public BSs, the decoding mode of the non-public BS over subcarriers, as well as the rate and power control. Numerical results show that the proposed design significantly improves the common uplink throughput as compared to benchmark schemes without such consideration.
\end{abstract}
\section{Introduction}
Recently, 5G non-public networks (NPNs) have attracted growing research interests to provide improved quality of service (QoS), higher security, and better privacy \cite{NPN} for supporting emerging applications such as industrial Internet of things (IIoT). Different from conventional 4G/5G public networks that are dominated by downlink traffic, 5G NPNs face significantly increased uplink traffic requirements \cite{uplink+}. Therefore, when the NPNs need to be deployed in conjunction with public networks \cite{3gpp}, the conventional uplink-downlink configuration for public networks (with downlink dominated) may not work well for NPNs \cite{configuration,UL-DL}.

In order to meet the significant uplink traffic requirements with scarce spectrum resources, the uplink-and-downlink spectrum sharing has emerged as a promising viable solution for NPN implementation, which allows the 5G NPNs to perform the uplink transmission over the time-frequency resources that are configured for downlink transmission in coexisting public networks,{\footnote{For instance, in the 3GPP Technical Specification 36.423, the uplink-and-downlink spectrum sharing function is enabled between 4G Evolved Universal Terrestrial Radio Access (E-UTRA) cells and 5G New Radio (NR) cells with overlapping coverage \cite{3gpp2}.}} thus enhancing the uplink throughput. Despite the benefit, however, the uplink-and-downlink spectrum sharing may cause harmful interference from the downlink transmission of public base stations (BSs) to the coexisting uplink non-public BS receivers. Such interference is particularly severe, as the transmit power at public BSs is generally much larger than that at users and the BS-to-BS channels are normally much stronger than the user-to-BS counterparts due to the BSs' relatively higher deployment locations. Therefore, the downlink-to-uplink asymmetric interference is a key challenge faced in the new uplink-and-downlink spectrum sharing scenario between NPNs and public networks.

In this paper, we propose a new adaptive asymmetric successive interference cancellation (SIC) approach to resolve the above problem for enhancing the uplink throughput of NPNs. In this approach, the uplink non-public BS receivers are enabled to have the capability of decoding the downlink signals transmitted from the public BSs and successively cancelling the resultant interference to facilitate the decoding of desirable uplink signals from non-public users. In particular, we focus our study on a basic scenario when an uplink non-public BS and a downlink public BS coexist in the same area, each communicating with multiple users via orthogonal frequency-division multiple access (OFDMA). To facilitate the adaptive asymmetric SIC, the uplink non-public and downlink public BSs cooperate in the subcarrier allocation and user scheduling, as well as the rate and power control over different subcarriers. Furthermore, depending on the channel conditions of different links, the non-public BS can adaptively choose the decoding mode at each subcarrier between SIC and treating interference as noise (TIN). Under this setup, we aim to maximize the common uplink throughput of all non-public users, while ensuring the downlink QoS requirement of public users such that the downlink throughput of each public user is above a certain threshold. Although the formulated throughput maximization problem is highly non-convex and difficult to be optimally solved, we propose an efficient algorithm to obtain a high-quality solution by using the techniques of alternating optimization and successive convex approximation (SCA). Numerical results show that the proposed adaptive asymmetric SIC design significantly improves the common uplink throughput as compared to benchmark schemes without the adaptive asymmetric SIC.

It is worth noting that the investigated uplink-and-downlink spectrum sharing between NPNs and public networks is different from the conventional spectrum sharing in cognitive radio (see, e.g., \cite{cr1,cr2,Rui}). In conventional cognitive radio, cognitive users try to access the licensed spectrum of the primary systems that generally belong to a different entity; in this case, the primary systems  are normally not aware of the existence of cognitive systems and thus cannot cooperate in helping the cognitive transmission. By contrast, in our considered uplink-and-downlink spectrum sharing, the public network is able to cooperate with the NPN by adjusting its downlink communication rate to facilitate the asymmetric SIC. Such cooperation is enabled, as the public network and NPN are interconnected via backhaul networks and may belong to the same entity (e.g., the same network operator) in practice.

\section{System Model}
In this paper, we consider an uplink-and-downlink spectrum sharing scenario when a public BS and a non-public BS coexist with overlapped coverage, and the public and non-public BSs share the same time-frequency resources for downlink and uplink transmission, respectively.{\footnote{For initial investigation, we focus on the interplay between one non-public BS and one public BS in this paper, and consider the interference from other non-public and public BSs as background noise. The proposed adaptive asymmetric SIC approach in this paper is also extendable to more general cases with multiple non-public and public BSs, e.g., by employing multiple antennas to help successively cancel the interference from multiple BSs.}} Suppose that there are $M^\text{UL}$ uplink users in the non-public cell, and $M^\text{DL}$ downlink users in the public cell, the sets of which are denoted as ${\cal M}^\text{UL}\!\triangleq\!\{1,...,M^\text{UL}\}$ and ${\cal M}^\text{DL}\!\triangleq\!\{1,...,M^\text{DL}\}$, respectively. We consider the OFDMA transmission for both the public and non-public BSs. Suppose that there are $N$ subcarriers for the OFDMA transmission, where ${\cal N}\!\triangleq\!\{1,...,N\}$ denotes the set of the $N$ subcarriers. Let $a^\text{UL}_{k,n}\!\in\!\{0,1\},k\!\in\!{\cal M}^\text{UL},n\!\in\!{\cal N},$ denote the subcarrier allocation indicator in the uplink non-public cell, where $a^\text{UL}_{k,n}\!=\!1$ indicates that subcarrier $n$ is allocated to non-public user $k$, and $a^\text{UL}_{k,n}\!=\!0$ otherwise. Similarly, let $a^\text{DL}_{l,n}\!\in\!\{0,1\},l\!\in\!{\cal M}^\text{DL},n\!\in\!{\cal N},$ denote the subcarrier allocation indicator in the downlink public cell, where $a^\text{DL}_{l,n}\!=\!1$ means that subcarrier $n$ is allocated to public user $l$, and $a^\text{DL}_{l,n}\!=\!0$ otherwise.
Notice that for OFDMA transmission, each subcarrier is allocated to at most one user in each cell. Therefore, we have
\begin{align}
&\sum\nolimits_{k\in{\cal M}^\text{UL}}a^\text{UL}_{k,n}\le1, \forall n\in{\cal N},\label{au}\\
&\sum\nolimits_{l\in{\cal M}^\text{DL}}a^\text{DL}_{l,n}\le1, \forall n\in{\cal N},
\label{al}
\end{align}
for uplink and downlink cells, respectively.
%
%\begin{figure}[t]
%\centering
%    \includegraphics[width=6cm]{model.eps}
%\caption{An uplink-and-downlink spectrum sharing scenario between a 5G non-public BS and a public BS.} \label{fig:model}
%\end{figure}

We consider quasi-static wireless channel models, where the wireless channels remain unchanged over the transmission block of our interest but may change over different blocks. Let ${f}_{k,n}$ denote the uplink channel power gain from non-public user $k\in{\cal M}^\text{UL}$ to the non-public BS at subcarrier $n\in{\cal N}$, ${\varphi}_{n}$ denote the channel power gain from the public BS to the non-public BS at subcarrier $n\in{\cal N}$, ${h}_{l,n}$ denote the downlink channel power gain from the public BS to public user $l\in{\cal M}^\text{DL}$ at subcarrier $n\in{\cal N}$, and ${g}_{k,l,n}$ denote the channel power gain from non-public user $k\in{\cal M}^\text{UL}$ to public user $l\in{\cal M}^\text{DL}$ at subcarrier $n\in{\cal N}$, respectively. It is assumed that there is a central controller that can perfectly obtain the global channel state information (CSI) for both the non-public and public cells. This assumption is made for obtaining the performance upper bound, which helps to gain essential insights in practice.

Next, we introduce the signal model. Let $s^\text{UL}_{k,n}$ denote the signal transmitted by non-public user $k\in{\cal M}^\text{UL}$ in the uplink and $s^\text{DL}_{l,n}$ that by the public BS for user $l\in{\cal M}^\text{DL}$ in the downlink at subcarrier $n$. Accordingly, the received signal at the non-public BS at subcarrier $n\in{\cal N}$ is expressed as
\begin{align}
y^\text{UL}_{n}=&~\sum\limits_{k\in {\cal M}^\text{UL}}\!a^\text{UL}_{k,n} s^\text{UL}_{k,n} \sqrt{f_{k,n}}\!+\!\sum_{l\in{\cal M}^\text{DL}}\!a^\text{DL}_{l,n}s^\text{DL}_{l,n}\sqrt{{\varphi}_{n}}\!+\!z^\text{UL}_{n},\label{ul}
\end{align}
where the first term $\sum\nolimits_{k\in {\cal M}^\text{UL}}a^\text{UL}_{k,n}s^\text{UL}_{k,n} \sqrt{f_{k,n}}$ at the right-hand-side (RHS) denotes the received signal from the non-public cell, the second term $\sum_{l\in{\cal M}^\text{DL}}a^\text{DL}_{l,n}s^\text{DL}_{l,n}\sqrt{{\varphi}_{n}}$ at the RHS denotes the inter-cell interference from the neighboring public cell, and $z^\text{UL}_{n}$ denotes the background noise (including the interference from other neighboring cells) that is assumed to be a circularly symmetric complex Gaussian (CSCG) random variable with zero mean and variance $\sigma^2$, i.e., $z^\text{UL}_{n}\sim\mathcal{CN}(0,\sigma^2)$.
On the other hand, the received signal at public user $l\in{\cal M}^\text{DL}$ at subcarrier $n\in{\cal N}$ is given by
\begin{align}
y^\text{DL}_{l,n}=~&\sqrt{{h}_{l,n}}a^\text{DL}_{l,n}s^\text{DL}_{l,n}+\sqrt{{h}_{l,n}}\sum\nolimits_{m\in{\cal M}^\text{DL},m\neq l}a^\text{DL}_{m,n}s^\text{DL}_{m,n}\notag\\
&+\sum\nolimits_{k\in{\cal M}^\text{UL}}a^\text{UL}_{k,n}s^\text{UL}_{k,n}\sqrt{{g}_{k,l,n}}+z^\text{DL}_{l,n},\label{dl}
\end{align}
where the first term $\sqrt{{h}_{l,n}}a^\text{DL}_{l,n}s^\text{DL}_{l,n}$ at the RHS denotes the desirable signal from the public BS, the second term $\sqrt{{h}_{l,n}}\sum\nolimits_{m\in{\cal M}^\text{DL},m\neq l}a^\text{DL}_{m,n}s^\text{DL}_{m,n}$ at the RHS denotes the intra-cell interference, the third term $\sum\nolimits_{k\in{\cal M}^\text{UL}}a^\text{UL}_{k,n}s^\text{UL}_{k,n}\sqrt{{g}_{k,l,n}}$ at the RHS denotes the inter-cell interference from the non-public cell, and $z^\text{DL}_{l,n}\sim\mathcal{CN}(0,\sigma^2)$ denotes the background noise.
We consider that the public BS and non-public users employ Gaussian signaling by setting $s^\text{UL}_{k,n}$'s and $s^\text{DL}_{l,n}$'s as independent CSCG random variables with zero mean, where $\mathbb{E}[|s^\text{UL}_{k,n}|^2]=p^\text{UL}_{k,n}$ and $\mathbb{E}[|s^\text{DL}_{l,n}|^2]=p^\text{DL}_{l,n}$ denote the transmit power of non-public user $k\in{\cal M}^\text{UL}$ and that of the public BS for user $l\in{\cal M}^\text{DL}$ at subcarrier $n\in{\cal N}$, respectively. Let $P^\text{DL}$ and $P_k^\text{UL},k\in{\mathcal M}^\text{UL}$ denote the maximum transmit power budgets by the public BS and each non-public user $k$, respectively. Then we have
\begin{align}
&\sum\nolimits_{n\in{\cal N}}\sum\nolimits_{l\in{\cal M}^\text{DL}} p^\text{DL}_{l,n}\leq P^\text{DL},\label{dl-power}\\
&\sum\nolimits_{n\in{\cal N}}p^\text{UL}_{k,n}\leq P_k^\text{UL}, \forall k\in{\cal M}^\text{UL}.\label{ul-power}
\end{align}

\subsection{Achievable Rates of Non-public Users with Asymmetric SIC}
In this subsection, we consider the achievable rates at the non-public users. In particular, we propose the adaptive asymmetric SIC approach, in which the non-public BS can adaptively switch between the SIC and the conventional TIN modes over each subcarrier to decode the information of non-public users. Let $\tau_{n}\!\in\!\{0,1\}$ denote the receiver mode indicator for the non-public BS to decode the message of non-public user $k\in{\cal M}^\text{UL}$ at subcarrier $n\!\in\!{\cal N}$, where $\tau_{n}\!=\!1$ corresponds to that the non-public BS adopts SIC for decoding; and $\tau_{n}\!=\!0$ means that TIN is employed.

First, we consider the asymmetric SIC at the non-public BS, in which the non-public BS first decodes the signals from the public BS and then cancels the resultant interference before decoding the desirable signals from non-public users.{\footnote{To facilitate the SIC at the non-public BS in practice, the public BS is enabled to share the employed modulation and coding schemes to the non-public BS via backhaul links.}} In order to facilitate the decoding, the public BS adopts the adaptive-rate transmission, by setting the communication rate as $r^\text{DL}_{l,n}$ for public user $l\in{\mathcal M}^\text{DL}$ at subcarrier $n\in {\mathcal N}$, which is a variable to be optimized. In this case, the non-public BS first decodes the message from the public BS by treating the signal from the non-public user as noise. Based on \eqref{ul}, the achievable rate of the communication link from the public BS to the non-public BS at subcarrier $n$ is given by
\begin{align}
&R^\text{BS}_{n}(\{a^\text{DL}_{l,n}\},\{p^\text{DL}_{l,n}\},\{a^\text{UL}_{k,n}\},\{p^\text{UL}_{k,n}\})\notag\\
&~~~~~~~~~~~~~~~=\log_2\!\left(\!1\!+\!\frac{\sum_{l\in{\cal M}^\text{DL}}a^\text{DL}_{l,n}p^\text{DL}_{l,n}\varphi_n}{\sum_{k\in{\cal M}^\text{UL}}a^\text{UL}_{k,n}p^\text{UL}_{k,n}{f}_{k,n}\!+\!\sigma^2}\!\right).\notag
\end{align}
In order for the SIC to be feasible at subcarrier $n\in{\cal N}$, we need to ensure that the achievable rate $R_n^\text{BS}(\{a^\text{DL}_{l,n}\},\{p^\text{DL}_{l,n}\},\{a^\text{UL}_{k,n}\},\{p^\text{UL}_{k,n}\})$ must be no less than the communication rate. Therefore, we have
\begin{align}
\tau_{n}\!\sum\limits_{l\in {\cal M}^\text{DL}}\!r^\text{DL}_{l,n}\leq R^\text{BS}_{n}(\{a^\text{DL}_{l,n}\},\{p^\text{DL}_{l,n}\},\{a^\text{UL}_{k,n}\},\{p^\text{UL}_{k,n}\}),\forall n\in{\cal N}.\label{bs-rate}
\end{align}
Notice that at the left-hand-side (LHS) in \eqref{bs-rate}, $\tau_n$ means that this constraint only need to hold when the SIC mode is adopted (i.e., $\tau_n = 1$), and $\sum\nolimits_{l\in{\cal M}^\text{DL}}r^\text{DL}_{l,n}$ is used as the scheduled non-public user at subcarrier $n\in{\cal N}$ to be decided. After decoding the downlink signals from the public BS, the non-public BS then successively cancels the resultant interference. Therefore, based on \eqref{ul} (with the term $\sum_{l\in{\cal M}^\text{DL}}\!a^\text{DL}_{l,n}s^\text{DL}_{l,n}\sqrt{{\varphi}_{n}}$ cancelled) together with the fact that only one non-public user can be scheduled (see \eqref{au}) at each subcarrier, the achievable rate of non-public user $k\in{\cal M}^\text{UL}$ at the non-public BS at subcarrier $n\in{\cal N}$ is given by
\begin{align}
\notag R^\text{UL-SIC}_{k,n}(a^\text{UL}_{k,n},p^\text{UL}_{k,n})=\log_2\left(1+{a^\text{UL}_{k,n}p^\text{UL}_{k,n}{f}_{k,n}}/{\sigma^2}\right).
\end{align}

Next, we consider the case with TIN receiver at subcarrier $n$ (i.e., $\tau_n=0$). In this case, the non-public BS decodes the message of non-public users by treating the interference from the public BS as noise. As a result, based on \eqref{au} and \eqref{ul}, the achievable rate of non-public user $k\in{\cal M}^\text{UL}$ at the non-public BS at subcarrier $n\in{\cal N}$ is given by
\begin{align}
&R^\text{UL-TIN}_{k,n}(a^\text{UL}_{k,n},p^\text{UL}_{k,n},\{a^\text{DL}_{l,n}\},\{p^\text{DL}_{l,n}\})\notag\\
&~~~=\log_2\left(1\!+\!{a^\text{UL}_{k,n}p^\text{UL}_{k,n}{f}_{k,n}}/\left({\sum\nolimits_{l\in{\cal M}^\text{DL}}a^\text{DL}_{l,n}p^\text{DL}_{l,n}\varphi_n\!+\!\sigma^2}\right)\right).\notag
\end{align}

By combining the two cases with SIC and TIN receivers, the total uplink throughput of each non-public user $k\in{\cal M}^\text{UL}$ over all subcarriers is defined as
\begin{align}
&\sum\limits_{n\in{\cal N}}((1-\tau_{n})R_{k,n}^\text{UL-TIN}(a^\text{UL}_{k,n},p^\text{UL}_{k,n},\{a^\text{DL}_{l,n}\},\{p^\text{DL}_{l,n}\})\notag\\
&~~~~~~+\tau_{n}R_{k,n}^\text{UL-SIC}(a^\text{UL}_{k,n},p^\text{UL}_{k,n})).\notag
\end{align}

\subsection{Achievable Rates of Public Users}
In this subsection, we consider the transmission at the public users. Based on \eqref{dl}, the downlink public user suffers from the inter-cell interference from the uplink non-public users. Note that it follows from \eqref{al} that only one public user is scheduled at each subcarrier. Therefore, by practically considering that no SIC is implementable at public user receivers, the achievable rate for public user $l\in{\cal M}^\text{DL}$ at subcarrier $n\in{\cal N}$ is given by
\begin{align}
&R^\text{DL}_{l,n}(a^\text{DL}_{l,n},p^\text{DL}_{l,n},\{a^\text{UL}_{k,n}\},\{p^\text{UL}_{k,n}\})\notag\\
&~=\log_2\left(1\!+\!{a^\text{DL}_{l,n}p^\text{DL}_{l,n}{h}_{l,n}}/\left({\sum\nolimits_{k\in{\cal M}^\text{UL}}a^\text{UL}_{k,n}p^\text{UL}_{k,n}{g}_{k,l,n}\!+\!\sigma^2}\right)\right).\notag
\end{align}
In order for each public user $l\in {\mathcal M}^\text{DL}$ to successfully decode the messages from the public BS, the communication rate $r^\text{DL}_{l,n}$ should not exceed the achievable rate $R_{l,n}^\text{DL}(a^\text{DL}_{l,n},p^\text{DL}_{l,n},\{a^\text{UL}_{k,n}\},\{p^\text{UL}_{k,n}\})$ of the corresponding link. Therefore, we have the following rate constraint for $r_{l,n}^\text{DL}$ besides that in \eqref{bs-rate}:
\begin{align}
r^\text{DL}_{l,n}\leq R^\text{DL}_{l,n}(a^\text{DL}_{l,n},p^\text{DL}_{l,n},\{a^\text{UL}_{k,n}\},\{p^\text{UL}_{k,n}\}),\forall l\in{\cal M}^\text{DL},n\in{\cal N}.\label{dl-rate}
\end{align}
Notice that to properly control the inter-cell interference at the public users, we optimize the pairing of the served public and non-public users at each subcarrier $n$ by deciding $\{a^\text{UL}_{k,n}\}$ and $\{a^\text{DL}_{l,n}\}$, as will be shown later.

\subsection{Problem Formulation}
In this paper, we aim to maximize the uplink throughput of non-public users in a fair manner, while ensuring the QoS requirements of public users. In particular, our objective is to maximize the common uplink throughput (i.e., $\min_{k\in{\cal M}^\text{UL}}\!\!\sum\nolimits_{n\in{\cal N}}((1-\tau_{n})R_{k,n}^\text{UL-TIN}(a^\text{UL}_{k,n},p^\text{UL}_{k,n},\{a^\text{DL}_{l,n}\},\{p^\text{DL}_{l,n}\})
+\!\tau_{n}R_{k,n}^\text{UL-SIC}(a^\text{UL}_{k,n},p^\text{UL}_{k,n}))$) of non-public users, by jointly optimizing the subcarrier allocation and user scheduling $\{a^\text{UL}_{k,n}\}$ and $\{a^\text{DL}_{l,n}\}$ for both non-public and public BSs, the decoding mode $\{\tau_{n}\}$ of the non-public BS over each subcarrier, as well as the uplink and downlink power control $\{p^\text{UL}_{k,n}\}$ and $\{p^\text{DL}_{l,n}\}$ and the downlink rate control $\{r^\text{DL}_{l,n}\}$ over subcarriers, subject to the subcarrier and power constraints in \eqref{au}, \eqref{al}, \eqref{dl-power}, and \eqref{ul-power} for both public and non-public cells and the minimum rate requirement $\Gamma_\text{min}$ for each public user. Therefore, the common uplink throughput maximization problem is formulated as
\begin{align}
\!&\!\textrm{(P1):}\!\!\!\!\!\!\mathop\mathtt{max}_{\substack{\{a^\text{UL}_{k,n}\},\{a^\text{DL}_{l,n}\},\{\tau_{n}\}\\\{p^\text{UL}_{k,n}\},\{p^\text{DL}_{l,n}\},\{r^\text{DL}_{l,n}\}}}
\!\min_{k\in{\cal M}^\text{UL}}\!\sum\limits_{n\in{\cal N}}((1\!-\!\tau_{n})R_{k,n}^\text{UL-TIN}\!+\!\tau_{n}R_{k,n}^\text{UL-SIC})\notag\\
&{\mathtt{s.t.}}~\sum\nolimits_{n\in{\cal N}}r^\text{DL}_{l,n}\geq \Gamma_\text{min},\forall l\in{\cal M}^\text{DL}\label{dl-threshold}\\
&~~~~~a^\text{UL}_{k,n}\!\in\!\{0,1\},a^\text{DL}_{l,n}\!\in\!\{0,1\},\!\forall k\!\in\!{\cal M}^\text{UL}\!\!, l\!\in\!{\cal M}^\text{DL}\!\!,n\!\in\!{\cal N}\!\label{subcarrier1}\!\\
&~~~~~\sum\nolimits_{k\in{\cal M}^\text{UL}}a^\text{UL}_{k,n}\le1, \sum\nolimits_{l\in{\cal M}^\text{DL}}a^\text{DL}_{l,n}\le1, \forall n\in{\cal N}\label{subcarrier2}\\
&~~~~~~\tau_{n}\in\{0,1\},\forall n\in{\cal N}\label{decodingmode}\\
&~~~~~\eqref{dl-power},\eqref{ul-power},\eqref{bs-rate},\eqref{dl-rate}.\notag
\end{align}
where \eqref{dl-power} and \eqref{ul-power} are power constraints for downlink and uplink transmission, \eqref{bs-rate}, \eqref{dl-rate}, and \eqref{dl-threshold} are the constraints for the downlink communication rate, \eqref{subcarrier1} and \eqref{subcarrier2} are subcarrier constraints for both non-public and public BSs, \eqref{decodingmode} is the constraint for decoding mode selection at the non-public BS. It is easy to observe that problem (P1) is a mixed-integer non-convex problem, in which the variables $\{a^\text{UL}_{k,n}\}$ and $\{a^\text{DL}_{l,n}\}$ are coupled with $\{p^\text{UL}_{k,n}\}$ and $\{p^\text{DL}_{l,n}\}$, respectively. Therefore, problem (P1) is challenging to be solved optimally.

Before proceeding, we check the feasibility of problem (P1), i.e., whether there exists a strategy such that the QoS requirements in $\Gamma_\text{min}$ for public users are met. In particular, checking the feasibility of problem (P1) is equivalent to maximizing the common downlink throughput (i.e., $\min_{l\in{\cal M}^\text{DL}}\sum\nolimits_{n\in{\cal N}}R^\text{DL}_{l,n}(a^\text{DL}_{l,n},p^\text{DL}_{l,n},\{a^\text{UL}_{k,n}\},\{p^\text{UL}_{k,n}\})$) of public users, by jointly optimizing the subcarrier and power allocation $\{a^\text{DL}_{l,n}\}$ and $\{p^\text{DL}_{l,n}\}$ for the public BS, subject to the subcarrier and power constraints, i.e.,
\begin{align}
\!&\mathop\mathtt{max}_{\substack{\{a^\text{DL}_{l,n}\},\{p^\text{DL}_{l,n}\}}}
\min_{l\in{\cal M}^\text{DL}}\sum\limits_{n\in{\cal N}}R^\text{DL}_{l,n}(a^\text{DL}_{l,n},p^\text{DL}_{l,n},\{a^\text{UL}_{k,n}\},\{p^\text{UL}_{k,n}\})\label{fp}\\
&~~~~~{\mathtt{s.t.}}~a^\text{DL}_{l,n}\in\{0,1\},\forall l\in{\cal M}^\text{DL},n\in{\cal N}\notag\\
&~~~~~~~~~~\sum\nolimits_{l\in{\cal M}^\text{DL}}a^\text{DL}_{l,n}\le1, \forall n\in{\cal N}\notag\\
&~~~~~~~~~~\sum\nolimits_{n\in{\cal N}}\sum\nolimits_{l\in{\cal M}^\text{DL}} p^\text{DL}_{l,n}\leq P^\text{DL}.\notag
\end{align}
Suppose that $L^*$ is the optimal value achieved by problem \eqref{fp}. Then it is evident that if $L^*\geq\Gamma_\text{min}$, then problem (P1) is feasible; otherwise, it is infeasible. Notice that problem \eqref{fp} is a conventional resource allocation problem for rate maximization in OFDMA systems, which has been well studied in the literature (see, e.g., \cite{LateifOFDM} for a similar solution). Therefore, we focus on the case when problem (P1) is feasible.

\section{Proposed Solution to Problem (P1)}
In this section, we propose an efficient algorithm to obtain a high-quality solution to problem (P1). To deal with the binary constraints in problem (P1), we first relax the binary variables $\{a^\text{UL}_{k,n}\}$, $\{a^\text{DL}_{l,n}\}$, and $\{\tau_{n}\}$ into continuous variables as
\begin{align}
&0\le \tau_{n}\le 1,\forall n\in{\cal N}\label{relaxed-tau},\\
&0\!\le\!a^\text{UL}_{k,n}\!\le\!1,0\!\le\!a^\text{DL}_{l,n}\!\le\!1,\forall k\!\in\!{\cal M}^\text{UL},l\!\in\!{\cal M}^\text{DL},n\!\in\!{\cal N}\label{relaxed-a}.
\end{align}
Then, we introduce auxiliary variables $R$, $\{E^\text{UL}_{k,n}\}$, and $\{E^\text{DL}_{l,n}\}$, with $E^\text{UL}_{k,n}\!=\!a^\text{UL}_{k,n}p^\text{UL}_{k,n},{k\in{\cal M}^\text{UL}}$ and $E^\text{DL}_{l,n}\!=\!a^\text{DL}_{l,n}p^\text{DL}_{l,n},{l\in{\cal M}^\text{DL},n\in{\cal N}}$. Accordingly, we have $p^\text{UL}_{k,n}\!=\!{E^\text{UL}_{k,n}}/{a^\text{UL}_{k,n}}$ and $p^\text{DL}_{l,n}\!=\!{E^\text{DL}_{l,n}}/{a^\text{DL}_{l,n}},$ where we define $p^\text{UL}_{k,n}\!=\!0, {k\in{\cal M}^\text{UL},n\in{\cal N}},$ if either $E^\text{UL}_{k,n}\!=\!0$ or $a^\text{UL}_{k,n}\!=\!0$ holds, and $p^\text{DL}_{l,n}\!=\!0,{l\in{\cal M}^\text{DL},n\in{\cal N}},$ if either $E^\text{DL}_{l,n}\!=\!0$ or $a^\text{DL}_{l,n}\!=\!0$ holds. By substituting ${E^\text{UL}_{k,n}}\!=\!p^\text{UL}_{k,n}{a^\text{UL}_{k,n}},$ and ${E^\text{DL}_{l,n}}\!=\!p^\text{DL}_{l,n}{a^\text{DL}_{l,n}},$ problem (P1) can be relaxed as
\begin{align}
\textrm{(P2):}&\mathop\mathtt{max}_{{\{a^\text{UL}_{k,n}\},\{a^\text{DL}_{l,n}\},\{\tau_{n}\},\{E^\text{UL}_{k,n}\},\{E^\text{DL}_{l,n}\},\{r^\text{DL}_{l,n}\},R\ge0}}
~R\notag\\
{\mathtt{s.t.}}&\sum\nolimits_{n\in{\cal N}} E^\text{UL}_{k,n}\le P_k^\text{UL},\forall k\in{\cal M}^\text{UL}\label{relaxed-power-ul}\\
&\sum\nolimits_{n\in{\cal N}}\sum\nolimits_{l\in{\cal M}^\text{DL}} E^\text{DL}_{l,n}\leq P^\text{DL}\label{relaxed-power-dl}\\
&\tau_{n}\sum\nolimits_{l\in {\cal M}^\text{DL}}r^\text{DL}_{l,n}\leq\bar{R}^\text{BS}_{n}(\{E^\text{DL}_{l,n}\},\{E^\text{UL}_{k,n}\}),\forall n\in{\cal N}\label{relaxed-bs-rate}\\
&r^\text{DL}_{l,n}\le
\bar{R}^\text{DL}_{l,n}(E^\text{DL}_{l,n},\{E^\text{UL}_{k,n}\}),\forall l\in{\cal M}^\text{DL},n\in{\cal N}\label{relaxed-dl-rate}\\
&\sum\nolimits_{n\in{\cal N}}((1\!-\!\tau_{n})\bar{R}_{k,n}^\text{UL-TIN}(E^\text{UL}_{k,n},\{E^\text{DL}_{l,n}\})\notag\\
&~~~~~~~~~~~~~~+\tau_{n}\bar{R}_{k,n}^\text{UL-SIC}(E^\text{UL}_{k,n}))\ge R,\forall k\!\in\!{\cal M}^\text{UL},\label{relaxed-ul-rate}\\
&\eqref{dl-threshold},\eqref{subcarrier2},\eqref{relaxed-tau},\eqref{relaxed-a},\notag
\end{align}
where
\begin{align}
\!&\bar{R}^\text{BS}_{n}(\{E^\text{DL}_{l,n}\},\{E^\text{UL}_{k,n}\})\!=\!\log_2\!\left(\sum_{k\in{\cal M}^\text{UL}}\!\!\!\!\!E^\text{UL}_{k,n}{f}_{k,n}\!\!+\!\!\!\!{\sum_{l\in{\cal M}^\text{DL}}\!\!\!\!E^\text{DL}_{l,n}\varphi_n\!\!+\!\sigma^2}\!\right)\notag\\
\!&~~~~~~~~~~~~~-\log_2\left({\sum\nolimits_{k\in{\cal M}^\text{UL}}\!\!E^\text{UL}_{k,n}{f}_{k,n}\!+\!\sigma^2}\!\right)\!,\forall n\in{\cal N},\label{SCAbs}\\
\!&\bar{R}^\text{DL}_{l,n}(E^\text{DL}_{l,n},\{E^\text{UL}_{k,n}\})\!=\!\log_2\!\left(\sum\limits_{k\in{\cal M}^\text{UL}}\!\!\!\!E^\text{UL}_{k,n}{g}_{k,l,n}\!+\!{E^\text{DL}_{l,n}{h}_{l,n}\!+\!\sigma^2\!}\right)\notag\\
\!&-\!\log_2\left({\sum\nolimits_{k\in{\cal M}^\text{UL}}E^\text{UL}_{k,n}{g}_{k,l,n}\!+\!\sigma^2}\right),\forall l\!\in\!{\cal M}^\text{DL},n\!\in\!{\cal N},\label{SCAdl}\\
\!&\bar{R}^\text{UL-TIN}_{k,n}(E^\text{UL}_{k,n},\{E^\text{DL}_{l,n}\})\!=\!\log_2\!\left(\sum\limits_{l\in{\cal M}^\text{DL}}\!\!\!E^\text{DL}_{l,n}\varphi_n\!+\!{E^\text{UL}_{k,n}{f}_{k,n}}\!+\!\sigma^2\!\right)\notag\\
\!&~\left.-\log_2\left({\sum\nolimits_{l\in{\cal M}^\text{DL}}E^\text{DL}_{l,n}\varphi_n\!+\!\sigma^2}\right)\right),\forall k\!\in\!{\cal M}^\text{UL},n\!\in\!{\cal N},\label{SCAul}\\
\!&\bar{R}^\text{UL-SIC}_{k,n}(E^\text{UL}_{k,n})=\log_2\left(1+{E^\text{UL}_{k,n}{f}_{k,n}}/{\sigma^2}\right).
\end{align}

However, problem (P2) is still non-convex due to non-convex constraints in \eqref{relaxed-bs-rate}, \eqref{relaxed-dl-rate}, and \eqref{relaxed-ul-rate}. In the following, we propose an alternating-optimization-based algorithm to solve problem (P2) by optimizing the resource allocation (i.e., subcarrier allocation and user scheduling $\{a^\text{UL}_{k,n}\}$ and $\{a^\text{DL}_{l,n}\}$, power control $\{E^\text{UL}_{k,n}\}$ and $\{E^\text{DL}_{l,n}\}$), as well as rate control $\{r^\text{DL}_{l,n}\}$) and the decoding mode $\{\tau_n\}$ of the non-public BS over each subcarrier (i.e., with or without SIC) in an iterative manner.

%To be specific, we first optimize the resource allocation under given decoding mode of the non-public BS, i.e., jointly optimize the subcarrier allocation and user scheduling for both non-public and public BSs, as well as the uplink and downlink power control over subcarriers. Then, we optimize the decoding mode of the non-public BS over each subcarrier (i.e., with or without SIC) under given resource allocation.

\subsubsection{Resource Allocation}
Under any given decoding mode $\{\tau_{n}\}$, the resource allocation optimization problem is formulated as
\begin{align}
\textrm{(P2.1):}~&\mathop\mathtt{max}_{\substack{\{a^\text{UL}_{k,n}\},\{a^\text{DL}_{l,n}\},\{E^\text{UL}_{k,n}\},\{E^\text{DL}_{l,n}\},\{r^\text{DL}_{l,n}\},R\ge0}}
R\notag\\
{\mathtt{s.t.}}~&\eqref{dl-threshold},\eqref{subcarrier2},\eqref{relaxed-a},\eqref{relaxed-power-ul},\eqref{relaxed-power-dl},\eqref{relaxed-bs-rate},\eqref{relaxed-dl-rate},\eqref{relaxed-ul-rate}\notag.
 \end{align}
Note that the constraints in \eqref{relaxed-bs-rate}, \eqref{relaxed-dl-rate}, and \eqref{relaxed-ul-rate} are non-convex, as the second terms at the RHS in \eqref{SCAbs}, \eqref{SCAdl}, and \eqref{SCAul} are all concave. To tackle the non-convexity, we adopt the SCA technique to update $\{E^\text{UL}_{k,n}\}$ and $\{E^\text{DL}_{l,n}\}$ in an iterative manner by approximating problem (P2.1) into a convex problem at each iteration. Suppose that $\{E^{\text{UL}(i)}_{k,n}\}$ and $\{E^{\text{DL}(i)}_{l,n}\}$ are the local point in the $(i\!+\!1)$-th iteration. Note that any concave function is globally upper-bounded by its first-order Taylor expansion at any point. Thus, we have
\begin{align}
&\log_2\!\left({\sum\nolimits_{k\in{\cal M}^\text{UL}}\!\!\!E^\text{UL}_{k,n}{f}_{k,n}\!+\!\sigma^2}\!\right)\notag\\
&\leq\log_2\!\left({\sum\limits_{k\in{\cal M}^\text{UL}}\!\!\!\!E^{\text{UL}(i)}_{k,n}\!{f}_{k,n}\!+\!\sigma^2}\!\right)\!+\!\!\!{\sum\limits_{k\in{\cal M}^\text{UL}}}\frac{f_{k,n}(E^{\text{UL}}_{k,n}\!-\!E^{\text{UL}(i)}_{k,n})}{\ln2( E^{\text{UL}(i)}_{k,n}\!f_{k,n}\!+\!\sigma^2)}\notag\\
&\triangleq \hat{R}^\text{ub}_{n},\label{upperbound1}\\
&\log_2\!\left({\sum\nolimits_{k\in{\cal M}^\text{UL}}\!\!\!E^\text{UL}_{k,n}{g}_{k,l,n}\!+\!\sigma^2}\right)\notag\\
&\leq\log_2\!\left({\sum\limits_{k\in{\cal M}^\text{UL}}\!\!\!\!E^{\text{UL}(i)}_{k,n}\!{g}_{k,l,n}\!+\!\sigma^2}\!\right)\!+\!\!\!\!{\sum\limits_{k\in{\cal M}^\text{UL}}}\frac{{g}_{k,l,n}(E^{\text{UL}}_{k,n}\!-\!E^{\text{UL}(i)}_{k,n})}{\ln2( E^{\text{UL}(i)}_{k,n}\!{g}_{k,l,n}\!+\!\sigma^2)}\notag\\
&\triangleq \tilde{R}^\text{ub}_{l,n},\label{upperbound2}\\
&\log_2\!\left({\sum\nolimits_{l\in{\cal M}^\text{DL}}\!\!\!E^\text{DL}_{l,n}\varphi_n\!+\!\sigma^2}\right)\notag\\
&\leq\log_2\!\left({\sum\limits_{l\in{\cal M}^\text{DL}}\!\!\!E^{\text{DL}(i)}_{l,n}\!\varphi_n\!+\!\sigma^2}\!\right)\!+\!\!\!\!{\sum\limits_{l\in{\cal M}^\text{DL}}}\frac{\varphi_n(E^{\text{DL}}_{l,n}\!-\!E^{\text{DL}(i)}_{l,n})}{\ln2( E^{\text{DL}(i)}_{l,n}\!\varphi_n\!+\!\sigma^2)}\notag\\
&\triangleq \check{R}^\text{ub}_{n}.\label{upperbound3}
\end{align}

Under any given $\{E^{\text{UL}(i)}_{k,n}\}$ and $\{E^{\text{DL}(i)}_{l,n}\}$, by using the upper bounds in \eqref{upperbound1}, \eqref{upperbound2}, and \eqref{upperbound3}, problem (P2.1) can be approximated as the following problem:
\begin{align}
&\textrm{(P2.2):}~\mathop\mathtt{max}_{\{a^\text{UL}_{k,n}\},\{a^\text{DL}_{l,n}\},\{E^\text{UL}_{k,n}\},\{E^\text{DL}_{l,n}\},\{r^\text{DL}_{l,n}\},R\ge0}
R\notag\\
&{\mathtt{s.t.}}~\tau_{n}\!\!\sum\limits_{l\in {\cal M}^\text{DL}}\!\!r^\text{DL}_{l,n}\!
\leq\!\log_2\!\left(\sum_{k\in{\cal M}^\text{UL}}\!\!\!\!E^\text{UL}_{k,n}{f}_{k,n}\!+\!\!\!{\sum_{l\in{\cal M}^\text{DL}}\!\!\!\!E^\text{DL}_{l,n}\varphi_n\!\!+\!\sigma^2}\!\right)\notag\\
&~~~~~~~~~~~~~~~~~~~~~~~~~~~~~~~~~~~~~~~~~~~~~~~~~~~-\!\hat{R}_{k,n}^\text{ub},\forall n\in{\cal N}\notag\\
&~~~~~r^\text{DL}_{l,n}\leq\log_2\!\left(\sum\limits_{k\in{\cal M}^\text{UL}}\!\!\!\!E^\text{UL}_{k,n}{g}_{k,l,n}\!+\!{E^\text{DL}_{l,n}{h}_{l,n}\!+\!\sigma^2\!}\right)\notag\\
&~~~~~~~~~~~~~~~~~~~~~~~~~~~~~~~~~~~~~~~-\!\tilde{R}_{l,n}^\text{ub},\forall l\in{\cal M}^\text{DL},n\in{\cal N}\notag\\
&~~~\sum\limits_{n\in{\cal N}}((1\!-\!\tau_{n})\!\left(\!\log_2\!\left(\sum\limits_{l\in{\cal M}^\text{DL}}\!\!\!E^\text{DL}_{l,n}\varphi_n\!+\!{E^\text{UL}_{k,n}{f}_{k,n}}\!+\!\sigma^2\!\right)\!-\!\check{R}^\text{ub}_n\right)\notag\\
&~~~~~~~~~~~~~~~~~~~~~~~~~~~+\tau_{n}\bar{R}_{k,n}^\text{UL-SIC}(E^\text{UL}_{k,n}))\ge R,\forall k\in{\cal M}^\text{UL}\notag\\
&~~~~~\eqref{dl-threshold},\eqref{subcarrier2},\eqref{relaxed-a},\eqref{relaxed-power-ul},\eqref{relaxed-power-dl}\notag.
\end{align}
It is evident that problem (P2.2) is a convex optimization problem, which can thus be efficiently solved via standard convex optimization methods, such as CVX toolbox \cite{CVX}. In each iteration $i$, we need to solve the convex optimization problem (P2.2) under given local point $\{a^{\text{UL}(i-1)}_{k,n}\}$, $\{a^{\text{DL}(i-1)}_{l,n}\}$, $\{E^{\text{UL}(i-1)}_{k,n}\}$, $\{E^{\text{DL}(i-1)}_{l,n}\}$, $ \{r^{\text{DL}(i-1)}_{l,n}\}$, and $R^{(i-1)}$, for which the optimal solution is given as $\{a^{\text{UL}(i)}_{k,n}\}$, $\{a^{\text{DL}(i)}_{l,n}\}$, $\{E^{\text{UL}(i)}_{k,n}\}$, $\{E^{\text{DL}(i)}_{l,n}\}$, $ \{r^{\text{DL}(i)}_{l,n}\}$, and $R^{(i)}$, which will be used as the local point for the next iteration $i+1$. In addition, the optimal value of problem (P2.2) is an under-estimate of that of problem (P2.1). Therefore, the achieved objective value of problem (P2.1) is non-decreasing after each iteration. As the optimal value of problem (P2.1) is upper bounded, it is evident that the iteration leads to a converged solution to problem (P2.1).

\subsubsection{Decoding Mode Optimization}
Under given resource allocation $\{a^\text{UL}_{k,n}\}$, $\{a^\text{DL}_{l,n}\}$, $\{E^\text{UL}_{k,n}\}$, $\{E^\text{DL}_{l,n}\}$, and $\{r^\text{DL}_{l,n}\}$, we optimize the decoding mode $\{\tau_{n}\}$, for which the optimization problem is expressed as
\begin{align}
\textrm{(P2.3):}~&\mathop\mathtt{max}_{\{\tau_{n}\},R\ge0}
R\notag\\
{\mathtt{s.t.}}~&\eqref{relaxed-tau},\eqref{relaxed-bs-rate},\eqref{relaxed-ul-rate}.\notag
%\eqref{relaxed-dl-rate}
\end{align}
Note that problem (P2.3) is a standard linear program (LP), which can thus be solved optimally via efficient convex optimization methods, such as CVX toolbox \cite{CVX}.

\subsubsection{Complete Algorithm for Solving (P1)}
By combining the solutions above, we solve problem (P2) by optimizing the resource allocation by solving problem (P2.1), and optimizing the decoding mode by solving problem (P2.3) in an alternative manner, respectively. Notice that for each iteration of variables update, the objective value of problem (P2) is monotonically non-decreasing and finite. Therefore, the proposed algorithm eventually results in a converged solution to problem (P2).

Furthermore, with the solution to problem (P2) at hand, the subcarrier allocation and user scheduling $\{a^\text{UL}_{k,n}\}$ and $\{a^\text{DL}_{l,n}\}$ as well as decoding mode $\{\tau_{n}\}$ need to be rounded as binary variables $\{a^{\text{UL}*}_{k,n}\}$, $\{a^{\text{DL}*}_{l,n}\}$, and $\{\tau^*_{n}\}$. Then we resolve the resource allocation optimization problem under given $\{a^{\text{UL}*}_{k,n}\}$, $\{a^{\text{DL}*}_{l,n}\}$, and $\{\tau^*_{n}\}$ to obtain the corresponding value of $R^*$ via alternating optimization and SCA similarly as for problem (P2.1). As a result, the approximate solution to problem (P1) is finally obtained.
\section{Numerical Results}
This section provides numerical results to validate the performance of our proposed adaptive asymmetric SIC design, as compared with the following two benchmark schemes.
\begin{itemize}
\item {\it SIC}: The non-public BS adopts SIC to decode messages of non-public users. This scheme corresponds to solving problem (P1) under fixed decoding mode with $\tau_{n}=1,\forall n\!\in\!{\cal N}$.
\item {\it TIN}: The non-public BS adopts TIN to decode messages of non-public users regarding the signal from the public BS as noise. This scheme corresponds to solving problem (P1) under fixed decoding mode with $\tau_{n}=0, \forall n\in{\cal N}$.
\end{itemize}
\begin{figure}[t]
\centering
    \includegraphics[width=6.6cm]{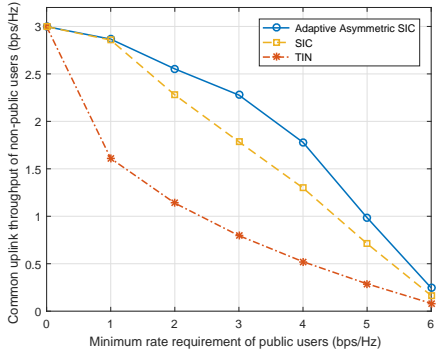}
\caption{Common uplink throughput versus the minimum rate requirement $\Gamma_\text{min}$ of public users.}
\label{Fig:Rmin}
\end{figure}

In the simulation, we set the locations of public and non-public BSs as $(0,0)$ and $(0,100~ \text{m})$, respectively, and the coverage area of each BS corresponds to a circle with a radius of $100$ m. The locations of public and non-public users are randomly generated in their respectively associated BS's coverage area. The wireless channels follow Rayleigh fading, specified by ${h}_{l,n}=|\theta_0(\bar{\Theta}_{l,n}/\Theta_0)^{-\xi}\bar{h}_{l,n}|^2$, ${\varphi}_{n}=|\theta_0(\tilde{\Theta}_{n}/\Theta_0)^{-\xi}\bar{\varphi}_{n}|^2$, ${f}_{k,n}=|\theta_0(\hat{\Theta}_{k,n}/\Theta_0)^{-\xi}\bar{f}_{k,n}|^2$, and ${g}_{k,l,n}=|\theta_0(\check{\Theta}_{k,l,n}/\Theta_0)^{-\xi}\bar{g}_{k,l,n}|^2$, where $\bar{h}_{l,n}$'s, $\bar{\varphi}_{n}$'s, $\bar{f}_{k,n}$'s, and $\bar{g}_{k,l,n}$'s are modeled as independent and identically distributed (i.i.d.) CSCG random variables with zero mean and unit variance, $\theta_0=-60$ dB corresponds to the path loss at the reference distance of $\Theta_0=10$ m, $\bar{\Theta}_{l,n}$, $\tilde{\Theta}_{n}$, $\hat{\Theta}_{k,n}$, and $\check{\Theta}_{k,l,n}$ denote the corresponding distance from the transmitter to the receiver, and $\xi=3$ is the pathloss exponent. The maximum transmit power of the public BS is set as $P^\text{DL}=40$ dBm, the noise power is set as $\sigma^2=-50$ dBm, and the maximum transmit power of non-public users is set as $P_k^\text{UL}=P_\text{max}=30$ dBm$,\forall k\in{\cal M}^\text{UL}$. We consider the case with $M_D=M_U=20$ and $N=100$. The bandwidth of each subcarrier is normalized. We conduct Monte Carlo simulations to show the average performance over 200 randomly realizations.

Fig. \ref{Fig:Rmin} shows the common throughput of non-public users versus the minimum rate requirement $\Gamma_\text{min}$. It is observed that as $\Gamma_\text{min}$ increases, the common uplink throughput of non-public users achieved by all the three schemes decreases, and the proposed adaptive-asymmetric-SIC approach significantly outperforms the other two benchmark schemes, thanks to the adaptive decoding mode selection between SIC and TIN over each subcarrier depending on the respective channel conditions. It is also observed that when $R_\text{min}$ is relatively small ($R_\text{min}\leq1$), the SIC scheme achieves a similar performance of the proposed adaptive-asymmetric-SIC approach, since the achievable rate of the link from the public BS to the non-public BS is generally higher than the communication rate, such that SIC is preferred over each subcarrier.

\begin{figure}[t]
\centering
    \includegraphics[width=6.6cm]{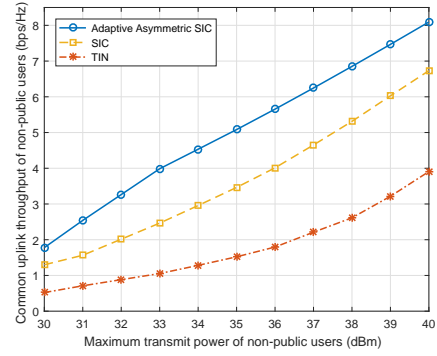}
\caption{Common uplink throughput versus the maximum transmit power $P_\text{max}$ of non-public users.}\label{Fig:power}
\end{figure}

Fig. \ref{Fig:power} shows the common uplink throughput versus the maximum transmit power of non-public users $P_\text{max}$, in which the minimum downlink throughput threshold is set as $\Gamma_\text{min}=4$ bps/Hz. It is observed that as the maximum transmit power $P_\text{max}$ increases, the common uplink throughput of non-public users achieved by all the three schemes increases considerably. It is also observed that the proposed adaptive-asymmetric-SIC approach significantly outperforms the other benchmark schemes without adaptive SIC, which validates the effectiveness of our proposed adaptive approach.

\section{Conclusion}\label{sec:V}
This paper studied a basic uplink-and-downlink spectrum sharing scenario when an uplink non-public BS shares the downlink time-frequency resources of a coexisting downlink public BS for uplink transmission. To deal with the severe co-channel interference from the downlink public BS to the coexisting uplink non-public BS, we proposed an adaptive asymmetric SIC approach at the non-public BS so as to enhance the uplink throughput. Under this setup, we maximized the common uplink throughput of all non-public users, under the condition that the downlink throughput of each public user is above a certain threshold. How to extend the proposed adaptive asymmetric SIC design to the scenarios with multiple non-public and public BSs is an interesting direction to be pursued in future research.


\begin{thebibliography}{1}
\bibliographystyle{IEEEbib}
\bibitem{NPN}
5G-ACIA, ``5G non-public networks for industrial scenarios (white paper),'' accessed on
May 25, 2020. [Online]. Available: {\url{https://www.5g-acia.org/publications/5g-non-public-networks-for-industrial-scenarios-white-paper}}
%\bibitem{uplink+}
%J. Oueis and E. C. Strinati, ``Uplink traffic in future mobile networks: Pulling the alarm,'' in {\it{Proc. CROWNCOM}}, Grenoble, 2016, pp. 583--593.
\bibitem{uplink+}
K. Au, L. Zhang, H. Nikopour, E. Yi, A. Bayesteh, U. Vilaipornsawai, J. Ma, and P. Zhu, ``Uplink contention based SCMA for 5G radio access,'' in {\it{Proc. IEEE GLOBECOM Workshops}}, Austin, 2014, pp. 900--905.


\bibitem{3gpp}
3GPP-TS-23.501, ``Study on system architecture for the 5G System (5GS),'' 3GPP Tech. Spec., 2017. [Online]. Available: {\url{http://www.3gpp.org/dynareport/23501.htm}}
\bibitem{configuration}
Z. Shen, A. Khoryaev, E. Eriksson, and X. Pan, ``Dynamic uplink-downlink configuration and interference management in TD-LTE,'' {\it IEEE Commun. Mag.}, vol. 50, no. 11, pp. 51--59, Nov. 2012.
\bibitem{UL-DL}
F. Boccardi, J. Andrews, H. Elshaer, M. Dohler, S. Parkvall, P. Popovski, and S. Singh, ``Why to decouple the uplink and downlink in cellular networks and how to do it,'' {\it IEEE Commun. Mag.}, vol. 54, no. 3, pp. 110--117, Mar. 2016.
%\bibitem{asy}
%C. Li, J. Wang, F. Zheng, J. M. Cioffi, and L. Yang, ``Overhearing-based cooperation for two-cell network with asymmetric uplink-downlink traffics,'' {\it IEEE Trans. Signal Inf. Process. Netw.}, vol. 2, no. 3, pp. 350--361, Sep. 2016.
%\bibitem{uplinkband}
%3GPP-TR-37.872, ``Supplementary uplink (SUL) and LTE-NR co-existence,'' 3GPP Tech. Rep., 2019. [Online]. Available: {\url{http://www.3gpp.org/dynareport/37872.htm}}

\bibitem{3gpp2}
3GPP-TS-36.423, ``Evolved universal terrestrial radio access network (E-UTRAN); X2 application protocol (X2AP),'' 3GPP Tech. Spec., 2020. [Online]. Available: {\url{http://www.3gpp.org/dynareport/36423.htm}}
\bibitem{cr1}
A. Goldsmith, S. A. Jafar, I. Maric, and S. Srinivasa, ``Breaking spectrum gridlock with cognitive radios: An information theoretic perspective,'' {\it{Proc. IEEE}}, vol. 97, no. 5, pp. 894--914, May 2009.
\bibitem{cr2}
I. F. Akyildiz, W. Lee, M. C. Vuran, and S. Mohanty, ``A survey on spectrum management in cognitive radio networks,'' {\it{IEEE Commun. Mag.}}, vol. 46, no. 4, pp. 40--48, Apr. 2008.
\bibitem{Rui}
R. Zhang and Y. Liang, ``Exploiting multi-antennas for opportunistic spectrum sharing in cognitive radio networks,'' {\it{IEEE J. Sel. Top. Signal Process.}}, vol. 2, no. 1, pp. 88--102, Feb. 2008.
%\bibitem{Niyato}
%D. Niyato and E. Hossain, ``Competitive pricing for spectrum sharing in cognitive radio networks: Dynamic game, inefficiency of nash equilibrium, and collusion,'' {\it{IEEE J. Sel. Areas Commun.}}, vol. 26, no. 1, pp. 192--202, Jan. 2008.
%\bibitem{ss-cr-survey}
%F. Hu, B. Chen, and K. Zhu, ``Full spectrum sharing in cognitive radio networks toward 5G: A survey,'' {\it{IEEE Access}}, vol. 6, pp. 15754--15776, Feb. 2018.

%\bibitem{Rui2}
%R. Zhang and Y. Liang, ``Investigation on multiuser diversity in spectrum sharing based cognitive radio networks,'' {\it{IEEE Commun. Lett.}}, vol. 14, no. 2, pp. 133--135, Feb. 2010.

%\bibitem{noma}
%W. Mei and R. Zhang, ``Uplink cooperative NOMA for cellular-connected UAV," {\it{IEEE J. Sel. Top. Signal Process.}}, vol. 13, no. 3, pp. 644--656, Jun. 2019.

%\bibitem{4}
%Y. Shi, E. Alsusa, A. Ebrahim, and M. W. Baidas, ``Uplink performance enhancement through adaptive multi-association and decoupling in UHF-mmWave hybrid networks,'' {\it IEEE Trans. Veh. Technol.}, vol. 68, no. 10, pp. 9735--9746, Oct. 2019.

%\bibitem{dude1}
%F. Boccardi, R. W. Heath, A. Lozano, T. L. Marzetta, and P. Popovski, ``Five disruptive technology directions for 5G,'' {\it{IEEE Commun. Mag.}}, vol. 52, no. 2, pp. 74--80, Feb. 2014.
%\bibitem{2}
%K. Sun, J. Wu, W. Huang, H. Zhang, H. Hsieh, and V. C. M. Leung, ``Uplink performance improvement for downlink-uplink decoupled HetNets with non-uniform user distribution,'' {\it IEEE Trans. Veh. Technol.}, vol. 69, no. 7, pp. 7518--7530, Jul. 2020.
%\bibitem{5}
%L. Zhang, W. Nie, G. Feng, F. Zheng, and S. Qin, ``Uplink performance improvement by decoupling uplink/downlink access in HetNets,'' {\it IEEE Trans. Veh. Technol.}, vol. 66, no. 8, pp. 6862--6876, Aug. 2017.



%\bibitem{hetnet}
%J. G. Andrews, H. Claussen, M. Dohler, S. Rangan, and M. C. Reed, ``Femtocells: Past, present, and future,'' {\it{IEEE J. Sel. Areas Commun.}}, vol. 30, no. 3, pp. 497--508, Apr. 2012.
%\bibitem{spectrum}
%Z. Sattar, J. V. C. Evangelista, G. Kaddoum, and N. Batani, ``Spectral efficiency analysis of the decoupled access for downlink and uplink in two-tier network,'' {\it{IEEE Trans. Veh. Technol.}}, vol. 68, no. 5, pp. 4871--4883, May 2019.
%\bibitem{dude2}
%H. Elshaer, F. Boccardi, M. Dohler, and R. Irmer, ``Downlink and uplink decoupling: A disruptive architectural design for 5G networks,'' in {\it{Proc. IEEE GLOBECOM}}, Austin, 2014, pp. 1798--1803.
%\bibitem{yuquan}
%Q. Yu, H. Zhou, J. Chen, J. Jing, J. J. Zhao, B. Qian, and J. Wang, ``A fully-decoupled RAN architecture for 6G inspired by neurotransmission,'' {\it {J. Commun. Inf. Netw.}}, vol. 4, no. 4, pp. 15--23, Dec. 2019.
%\bibitem{twotier}
%M. Bacha, Y. Wu, and B. Clerckx, ``Downlink and uplink decoupling in two-tier heterogeneous networks with multi-antenna base stations,'' {\it{IEEE Trans. Wireless Commun.}}, vol. 16, no. 5, pp. 2760--2775, May 2017.
%\bibitem{coverageanalysis}
%F. Muhammad, Z. H. Abbas, G. Abbas, and L. Jiao, ``Decoupled downlink-uplink coverage analysis with interference management for enriched heterogeneous cellular networks,'' {\it{IEEE Access}}, vol. 4, pp. 6250--6260, 2016.
%\bibitem{ULspe}
%R. Li, K. Luo, T. Jiang, and S. Jin, ``Uplink spectral efficiency analysis of decoupled access in multiuser MIMO HetNets,'' {\it{IEEE Trans. Veh. Technol.}}, vol. 67, no. 5, pp. 4289--4302, May 2018.
%\bibitem{TMC}
%B. Lahad, M. Ibrahim, S. Lahoud, K. Khawam, and S. Martin, ``Joint modeling of TDD and decoupled uplink/downlink access in 5G HetNets with multiple small cells deployment,'' {\it{IEEE Trans. Mobile Comput.}}, 2020.
%\bibitem{analysis}
%A. I. Aravanis, O. Munoz, A. Pascual-Iserte, and J. Vidal, ``Analysis of downlink and uplink decoupling in dense cellular networks,"  in {\it{Proc. IEEE CAMAD}}, Toronto, 2016, pp. 219--224.
%\bibitem{analysis2}
%K. Smiljkovikj, P. Popovski, and L. Gavrilovska, ``Analysis of the decoupled access for downlink and uplink in wireless heterogeneous networks,'' {\it{IEEE Wireless Commun. Lett.}}, vol. 4, no. 2, pp. 173--176, Apr., 2015.
%\bibitem{cellassociation}
%H. Elshaer, M. N. Kulkarni, F. Boccardi, J. G. Andrews, and M. Dohler, ``Downlink and uplink cell association with traditional macrocells and millimeter wave small cells,'' {\it{IEEE Trans. Wireless Commun.}}, vol. 15, no. 9, pp. 6244--6258, Sept. 2016.
%\bibitem{utility}
%Y. Lin, W. Bao, W. Yu, and B. Liang, ``Optimizing user association and spectrum allocation in HetNets: A utility perspective,'' {\it{IEEE J. Sel. Areas Commun.}}, vol. 33, no. 6, pp. 1025--1039, Jun. 2015.
%\bibitem{backhaul}
%N. Sapountzis, T. Spyropoulos, N. Nikaein, and U. Salim, ``User association in HetNets: Impact of traffic differentiation and backhaul limitations,'' {\it{IEEE/ACM Trans. Netw.}}, vol. 25, no. 6, pp. 3396--3410, Dec. 2017.
%
%
%
%
%%%%
%
%
%
%
%
%
%
%
%
%
%\bibitem{11}
%M. Chen, W. Saad, and C. Yin, ``Echo state networks for self-organizing resource allocation in LTE-U with uplink-downlink decoupling,'' {\it{IEEE Trans. Wireless Commun.}}, vol. 16, no. 1, pp. 3--16, Jan. 2017.
%
%
%
%
%\bibitem{10}
%S. Singh, X. Zhang, and J. G. Andrews, ``Joint rate and SINR coverage analysis for decoupled uplink-downlink biased cell associations in HetNets,'' {\it{IEEE Trans. Wireless Commun.}}, vol. 14, no. 10, pp. 5360--5373, Oct. 2015.
%\bibitem{3}
%J. Zheng, J. Li, N. Wang, and X. Yang, ``Joint load balancing of downlink and uplink for eICIC in heterogeneous network,'' {\it IEEE Trans. Veh. Technol.}, vol. 66, no. 7, pp. 6388--6398, Jul. 2017.

%\bibitem{6}
%H. Elshaer, M. N. Kulkarni, F. Boccardi, J. G. Andrews, and M. Dohler, ``Downlink and uplink cell association with traditional macrocells and millimeter wave small cells,'' {\it IEEE Trans. Wireless Commun.}, vol. 15, no. 9, pp. 6244--6258, Sept. 2016.
%\bibitem{r1}
%L. Zhang, Y. Xin, Y. Liang, and H. Poor, ``Cognitive multiple access channels: Optimal power allocation for weighted sum rate maximization,'' {\it{IEEE Trans. Commun.}}, vol. 57, no. 9, pp. 2754--2762, Sep. 2009.
%\bibitem{r2}
%P. Cheng, G. Yu, Z. Zhang, H. H. Chen, and P. Qiu, ``On the achievable rate region of Gaussian cognitive multiple access channel,'' {\it{IEEE Commun. Lett.}}, vol. 11, no. 5, pp. 384--386, May 2007.


%\bibitem{r1}
%H. Lin, K. Ishibashi, W. Shin, and T. Fujii, ``Decentralized power allocation for secondary random access in cognitive radio networks with successive interference cancelation,'' in {\it{Proc. IEEE ICC}}, Kuala Lumpur, May 2016, pp. 1--6.
%\bibitem{r2}
%B. Maham, P. Popovski, and X. Zhou, ``Opportunistic interference cancelation and user selection in cognitive multiple access network,'' in {\it{Proc. IEEE SPAWC}}, Cesme, Jun. 2012, pp. 174--178.

%\bibitem{CVX}
%S. Boyd. \textit{EE364b Convex Optimization II, Course Notes}, accessed on
%Jan. 13, 2020. [Online] Available: \url{http://www.stanford.edu/class/ee364b/}
%[Online] Available: {\url{https://arxiv.org/abs/1008.3437}}
%\bibitem{Asymetric}
%Y. Zeng, J. Lyu, and R. Zhang, ``Cellular-connected UAV: Potential, challenges, and promising technologies,'' {\em IEEE Wireless Commun.}, vol. 26, no. 1, pp. 120--127, Feb. 2019.
%\bibitem{noma}
%Z. Ding, Y. Liu, J. Choi, Q. Sun, M. Elkashlan, C.-L. I, and H. V. Poor, ``Application of non-orthogonal multiple access in LTE and 5G networks,'' {\it{IEEE Commun. Mag.}}, vol. 55, no. 2, pp. 185--191, Feb. 2017.
%\bibitem{SIC11}
%M. Kamel, W. Hamouda, and A. Youssef, ``Uplink performance of NOMA-based combined HTC and MTC in ultradense networks,'' {\it{IEEE Internet Things J.}}, vol. 7, no. 8, pp. 7319--7333, Aug. 2020.
%\bibitem{buffer}
%R. Liu, P. Popovski, and G. Wang, ``Decoupled uplink and downlink in a wireless system with buffer-aided relaying,'' {\it{IEEE Trans. Commun.}}, vol. 65, no. 4, pp. 1507--1517, Apr. 2017.
%\bibitem{region}
%M. Charafeddine, A. Sezgin, Z. Han, and A. Paulraj, ``Achievable and crystallized rate regions of the interference channel with interference as noise'', {\it IEEE Trans. Wireless Commun.}, vol. 11, no. 3, pp. 1100--1111, Mar. 2012.
%\bibitem{ratesplitting}
%H. Yagi and H. V. Poor, ``Multi-level rate-splitting for synchronous and asynchronous interference channels,'' in {\it{Proc. IEEE ISIT}}, St. Petersburg, 2011, pp. 2080--2084.
%\bibitem{noma}
%M. Zeng, W. Hao, O. A. Dobre, and Z. Ding, ``Cooperative NOMA: State of the art, key techniques, and open challenges,'' {\it{IEEE Netw.}}, vol. 34, no. 5, pp. 205--211, Sept. 2020.
%\bibitem{noma2}
%W. Mei and R. Zhang, ``Cooperative NOMA for downlink asymmetric interference cancelation,'' {\it{IEEE Wireless Commun. Lett.}}£¬ vol. 9, no. 6, pp. 884--888, Jun. 2020.
%\bibitem{tel}
%\bibitem{LateifOFDM2}
%J. Liu, W. Chen, Z. Cao, and K. B. Letaief, ``Dynamic power and sub-carrier allocation for OFDMA-based wireless multicast systems,'' in {\it{Proc. IEEE ICC}}, Beijing, 2008, pp. 2607--2611.
\bibitem{LateifOFDM}
Y. J. Zhang and K. B. Letaief, ``Multiuser adaptive subcarrier-and-bit allocation with adaptive cell selection for OFDM systems,'' {\it{IEEE Trans. Wireless Commun.}}, vol. 3, no. 5, pp. 1566--1575, Sept. 2004.
\bibitem{boyd}
S. Boyd and L. Vandenberghe, {\em Convex optimization}, Cambridge, U.K.: Cambridge Univ. Press, Mar. 2004.

%M. N. Sial and  J. Ahmed, ``Analysis of K-tier 5G heterogeneous cellular network with dual-connectivity and uplink-downlink decoupled access,'' {\it{Telecommun Syst.}}, vol. 67, pp. 669--685, Apr. 2018.
\bibitem{CVX}
M. Grant and S. Boyd, {\it{CVX: MATLAB Software for Disciplined Convex Programming}}, 2016. [Online] Available: {\url{https://cvxr.com/cvx}}

\end{thebibliography}
\end{document}